\documentclass[letterpaper]{article}

\usepackage{emulateapj}
\usepackage{onecolfloat}
\usepackage{apjfonts}
\usepackage{amsmath}
\usepackage{natbib}
\usepackage{graphicx}

\newcommand{\SO}{XTE~J1550$-$564}
\newcommand{\LQ}{low frequency QPO}
\newcommand{\HQ}{high frequency QPO}
\newcommand{\HF}{high frequency}
\newcommand{\LF}{low frequency}

\begin{document}

\submitted{Accepted by the Astrophysical Journal}


\twocolumn[
\title{{X}-ray observations of {XTE J1550$-$564} during the decay of the 2000 
outburst - II. {T}iming }

\author{E. Kalemci\altaffilmark{1},
 J. A. Tomsick\altaffilmark{1},
 R. E. Rothschild\altaffilmark{1},
 K. Pottschmidt\altaffilmark{2},
 P. Kaaret\altaffilmark{3},
}

\altaffiltext{1}{Center for Astrophysics and Space Sciences, Code
0424, University of California at San Diego, La Jolla, CA,
92093-0424, USA}

\altaffiltext{2}{Institut f\"{u}r Astronomie und Astrophysik,
Astronomie, University of T\"{u}bingen, Waldh\"{a}user Strasse 64,
D-72076 T\"{u}bingen, Germany}

\altaffiltext{3}{Harvard-Smithsonian Center for Astrophysics, 60 Garden Street,
 Cambridge, CA, 02138, USA}


\begin{abstract}

     We investigate the timing behavior of \SO\ with the \emph{Rossi X-ray 
Timing Explorer (RXTE)} as the source made a transition to the hard state 
during the decay of the 2000 outburst. We detect a \HQ\ at 65 Hz in one 
observation with a fractional rms amplitude of 4.9\%. This is the first time 
that a \HQ\ has been detected in a transition to the hard state during outburst
decay.  We also observe \LQ s in the 0.36 -- 4.1 Hz range and rich aperiodic 
variability. The changes in the temporal properties  during the decay are very 
similar to the state transitions observed for other sources. We find a 
correlation between the \LF\ QPO and the break frequency in the continuum power
 spectra. We investigate the energy dependence of rms amplitudes of the QPOs. 
We compare these timing properties with those previously observed for \SO\ and 
other Galactic black hole candidates and discuss the implications for QPO 
models and black hole accretion.

\end{abstract}

\keywords{stars:individual (XTE J1550$-$564) -- black hole physics -- 
X-rays:stars}

] 


\section{Introduction}\label{sec:intro}
 
     The Soft X-ray Transient \SO\ was first detected  by the All-Sky Monitor 
\citep[ASM;][]{Levine96} on board the \emph{Rossi X-ray Timing Explorer (RXTE)}
 in 1998 September \citep{Smith98}. The power spectra for \SO\ showed strong 
aperiodic variability as well as  \LF\ (0.08 -- 18 Hz) 
\citep{Cui99,Remillard01} and \HF\ (100 -- 284 Hz) \citep{Remillard99,Homan00} 
quasi-periodic oscillations (QPOs) during some of the \emph{RXTE} observations.
 Previously detected \HQ s had low rms amplitudes ($\rm \sim 1\,\%$) and 
occurred mostly when the source was bright. The low frequency QPOs occurred for
 all of the spectral states. (The spectral states of the black hole binaries
are not rigorously defined. See \cite{Homan00} for a discussion of of these 
states in \SO .) The amplitudes of these \LQ s can be less than 
1\% in the soft state and as much as 17\% in the hard state. Detailed 
discussion of classification of these QPOs and their properties can be found 
in \cite{Homan00} and \cite{Remillard01}. The X-ray energy spectra 
usually show a soft component which can be modeled as blackbody emission over 
a range of temperatures from a geometrically thin, optically thick disk and a 
hard component modeled by a power-law which is thought to be the result of 
Compton up-scattering of soft photons by an energetic electron corona 
\citep{Sobczak99}. Both the X-ray spectra and the temporal properties are 
similar to the other Galactic black hole candidates (BHC), and the
recently measured mass function of $\rm 6.86 \pm 0.71 \; M_{\odot}$ confirms 
its black hole nature \citep{Orosz_at}.

     During the 1998 outburst, optical \citep{Orosz_iauc98} and radio 
\citep{CampbellW_iauc98} counterparts were identified. A superluminal ejection 
was discovered in the radio establishing the source as a microquasar, similar 
to GRO~J1655$-$40 and GRS~1915+105 \citep{Hannikainen00}. It became active in 
X-rays again in 2000 April \citep{Smith00}. The outburst reached its maximum 
ASM count rate at the end of 2000 April. The X-ray flux decayed over the next
few months and a transition to the hard state occurred. For more information 
about the X-ray spectrum and the radio properties of this 
outburst, see \cite{Tomsick_submitted} (hereafter paper I), \cite{Corbel01}, 
and \cite{Miller01}.

     In this paper, we used the PCA/\emph{RXTE} (Proportional Counter Array, 
see \cite{Bradt93} for a description of \emph{RXTE}) power spectra from 
observations of \SO\ made between 16 May and 3 June, during the decay of the 
2000 outburst. The detailed spectral analysis with \emph{RXTE} and 
\emph{Chandra} during the decay is covered by paper I. We searched for \HQ s 
and studied the evolution of the \LQ s through the decay. We also investigated
 the evolution of the continuum parameters of the power spectra during the 
transition, and compared the temporal properties to the previous observations 
and other BHCs.


\section{Observations and Analysis}\label{sec:obs}

     We used the standard \emph{RXTE} data analysis software FTOOLS 5.0 to 
extract light curves.  In the first 2 observations, the PCA data were 
accumulated in single bit mode  with $\rm 125\,\mu s$ time resolution for 
two energy bands, 2 -- 6 keV and 6 -- 15 keV. Above 15 keV, we used the event 
mode with $\rm8\,\mu s$ time resolution and with 32 energy channels. After the 
second observation, the data were accumulated in the event mode with 
$\rm 125\,\mu s$ time resolution and 64 energy channels. Typical integration 
times were 1 -- 2 ks per observation. The \SO\ light curve, and the exact 
integration times for our observations can be found in paper I. We used all of 
the Proportional Counter Units (PCU) that were on simultaneously. The 
observations were made after the loss of the propane layer for PCU~0, and we 
dealt with the additional background by assuming that PCU~0  and PCU~2 are 
identical and that the difference in the count rates between the two is the 
excess background from PCU~0. This excess background is added to the background
 obtained by using ``pcabackest''. It should be noted that the 
background is only used to convert the power spectra to rms normalization 
\citep{Berger94}. For each observation, we computed the power spectra using IDL
 programs developed at the University of T\"{u}bingen for three energy bands, 
2 -- 6 keV, 6 -- 15 keV, and 15 -- 67 keV and also for the summed band of 
2 -- 67 keV. Above 67 keV, the source is not significantly above the 
background. The power spectra were corrected for the dead time effects 
according to \cite{Zhang95} with a dead-time of $\rm 10\,\mu s$ per event. 

     First, we searched for high frequency QPOs by using short 1 second 
segments with a Nyquist frequency of 2048 Hz using the summed energy band. 
Except for observation 4 with its 65 Hz QPO, we did not detect any high
frequency QPO. The 95\% confidence upper limit on the rms amplitude is 
$\rm \sim 3\%$ before observation~6, and $\rm \sim 5\%$ after observation~6
for QPOs above 50 Hz. Then, using 128 second time segments, we investigated the
 \LQ s and the timing properties of the continuum  up to 256 Hz.


\section{Results}\label{sec:results}

\begin{figure}[b]
\vspace{0.2in}
\plotone{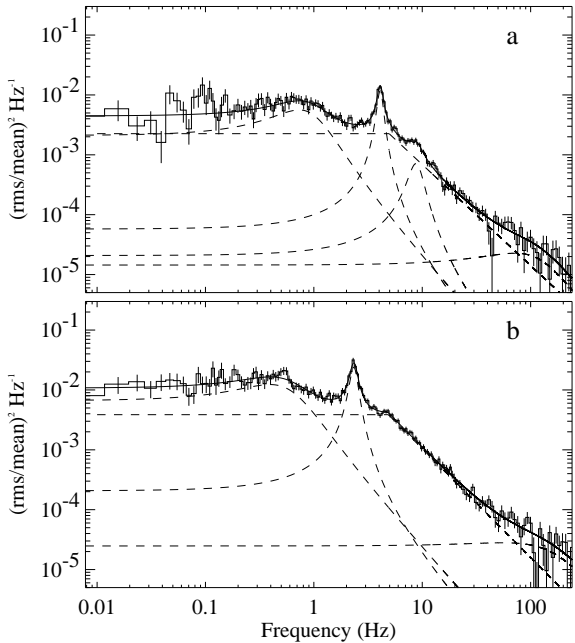}
\caption{\label{fig:ob1ps} 
Power spectrum and the best fit model for observations~1 (a) and 2 (b). For 
observation~1, the model is FT (broken power-law with the index of the 
power-law below the break fixed to zero) plus two narrow Lorentzians for the 
QPO and its first harmonic plus 2 broad  Lorentzians for the \LF\ and the \HF\ 
noise components. For observation~2 (and also 3), the second harmonic is 
not statistically significant and the model is FT + QPO + 2 broad Lorentzians.
The dashed lines represent individual components and the solid line is the 
overall fit to the data.
}
\end{figure}

     We used observation~1 (Fig.~\ref{fig:ob1ps}a) as a test case to establish 
the model for fitting the power spectra. We started with a broken power-law 
with the index of the power-law below the break fixed to zero (hereafter we 
refer to this model as flat top or FT) to fit the continuum, since this 
component is seen in all observations. To this, we added a Lorentzian at
  $\rm \, \sim \,$4 Hz to fit the \LQ. There were significant residuals both 
at the high and \LF\ ends of the spectrum, and we obtained a poor fit with a 
$\rm \chi^{2}$ of 483 with 150 degrees of freedom. Adding a broad low 
frequency bump (modeled by a Lorentzian) decreased the $\rm \chi^{2}$ 
to 235/147. The $\rm \chi^{2}$ dropped to 169/144 when another  broad 
Lorentzian was added to fit the continuum at high frequencies. Even after this,
 we observed excess power at around twice the QPO frequency (first harmonic) 
and added a narrow Lorentzian which reduced the $\rm \chi^{2}$ to 153/141. 
Fig.~\ref{fig:ob1ps}a shows the power spectrum of observation 1 and the full 
model, which is FT plus 4 Lorentzians (\HF\ bump, \LF\ bump, QPO and the 
harmonic). For the other observations, not all components are present at a 
significant level. (cf. Table~\ref{table:par}, Table~\ref{table:qpo} and the 
following discussion). 

\begin{figure}[t]
\centerline{\includegraphics[height=6in]{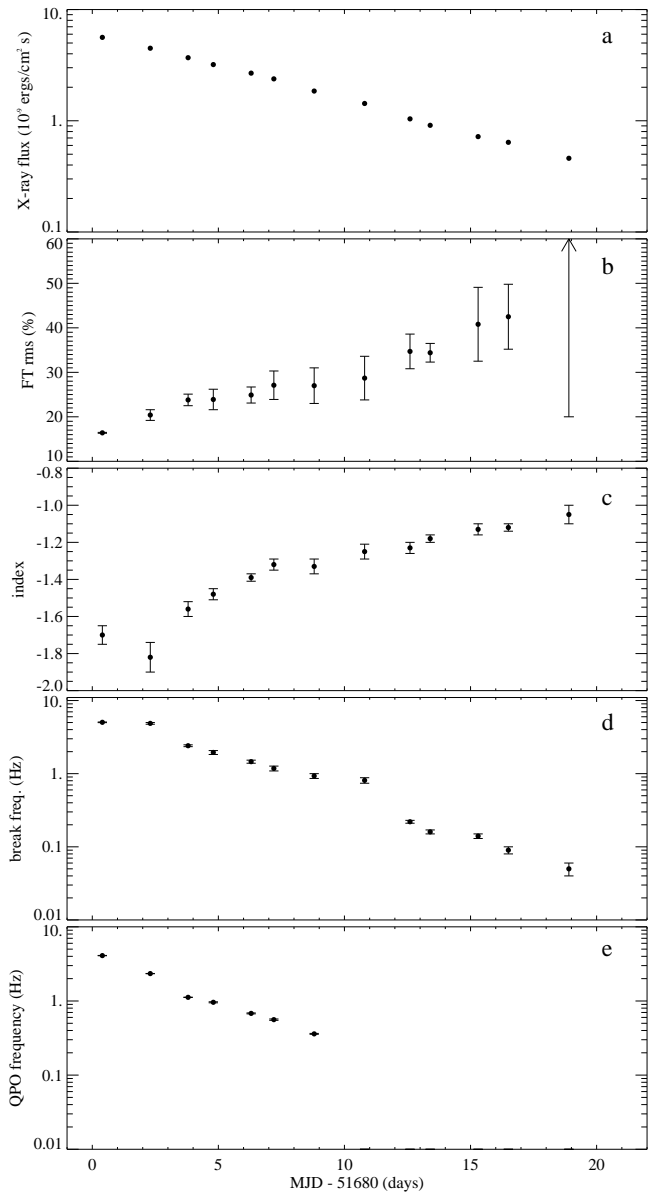}}
\vspace{0.5cm}
\caption{\label{fig:par}
The change in the parameters of the power spectra as the flux decreases and the
source makes a transition to the hard state. (a) is the 2.5 -- 20 keV X-ray 
flux, (b) is the rms amplitude of the FT component, (c) is the index of 
power-law after the break in FT, (d) is the break frequency, and (e) is the 
QPO frequency.}
\end{figure}

\begin{figure}[b]
\plotone{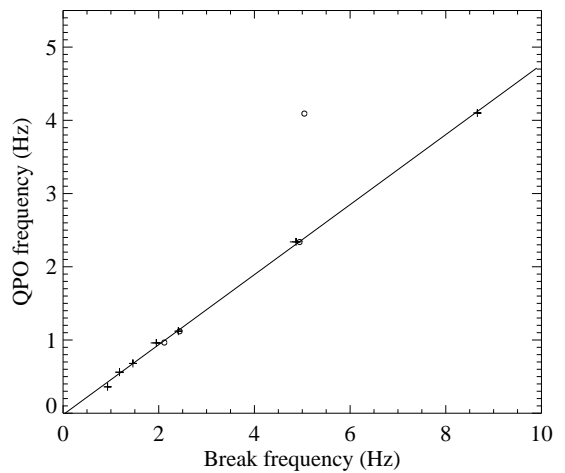}
\caption{\label{fig:vbvf}  
The correlation between the \LQ\ and the break frequency. Crosses indicate the
case when we limit the model to FT + 2 broad Lorentzians without the first 
harmonic and the circles indicate the case where we include a first harmonic
even if it is not statistically significant. The correlation breaks down in
observation~1 where a significant first harmonic is present. For the crosses,
the fit is a straight line with equation: 
$\rm \nu_{QPO} = (0.478 \pm 0.024) \;\nu_{break}\,-\,(0.020 \pm 0.006) \,Hz$ 
and the correlation coefficient is 0.9996.
}
\end{figure}

    A \LQ\ occurs for each observation between observation~1 and observation~7
 and the frequency of the QPO decreases from 4.09 Hz to 0.36 Hz as the overall
flux drops by a factor of 3 (see Table~\ref{table:par}, Table~\ref{table:qpo},
 and Fig.~\ref{fig:par}). The amplitude stays around 10\% and then decreases to
 7.1\% in observation~7 where it is last detected. The QPO is always below the 
break frequency. In observation~8, the 90\% confidence upper-limit for a 
QPO below the break frequency is 7.3\%.  The first harmonic is statistically 
significant only in the first observation and has an rms amplitude of 5.8\%. 
For observations 2,3 and 4, the 90\% confidence upper limits of the rms 
amplitude of the harmonic is between 1.8 -- 4.2\% (see Table~\ref{table:qpo}).
 The frequency of the first harmonic in observation~1 is slightly higher than 
twice the QPO frequency, and this may be due to the presence of a second 
harmonic, although adding another Lorentzian at the second harmonic frequency 
does not improve the fit. The QPO frequencies are correlated with the X-ray 
flux (Fig.~\ref{fig:par}) and the break frequencies (Fig.~\ref{fig:vbvf}). 
The quality-values ($Q$-values) are between 3 -- 7. We also 
investigated the energy dependence of the rms amplitude of the \LQ s. Low 
count rates allowed us to use only three energy bands: 2 -- 6, 6 -- 15, and 
15 -- 67 keV, and Fig.~\ref{fig:rmsan} shows the results. The rms amplitude 
shows a turnover in the first observation, an increase with energy in the 
second observation and a decrease with energy in the remaining observations.

   For the continuum parameters, the high frequency bump is
 required for the first three observations (See Fig.~\ref{fig:ob1ps} for 
observations~1 and 2). It peaks between $\rm \sim$ 40 -- 60 Hz and its 
amplitude is between 7.1\% and 11.8\%. The observation 3 power spectrum is 
very similar to that of observation~2. For observation~4 there is no high
 frequency bump, but a QPO is observed at 65 Hz.  After observation~4, adding 
a high frequency component does not improve the fit. The \LF\ noise, which 
peaks between 0.1 Hz and 0.8 Hz,  persists longer with an amplitude range of 
7.6\% to 18.8\%. It does not show a decreasing or increasing trend in time and 
disappears after observation~8 (see Table~\ref{table:par}). Therefore, 
starting from observation~9, the power spectrum can be represented with just a
 FT component. The amplitude of the FT component increases from 16\% in 
observation~1 to 43\% in observation~12 (Fig.~\ref{fig:par}b). The power-law 
index and the break frequency both show clear trends: the index 
(Fig.~\ref{fig:par}c) increases gradually while the break frequency 
(Fig.~\ref{fig:par}d.) decreases. All these changes are consistent with a 
transition to a hard state \citep{Vanderklis95}, and the energy spectrum also 
shows this transition (paper~I). 

\begin{figure}[t]
\plotone{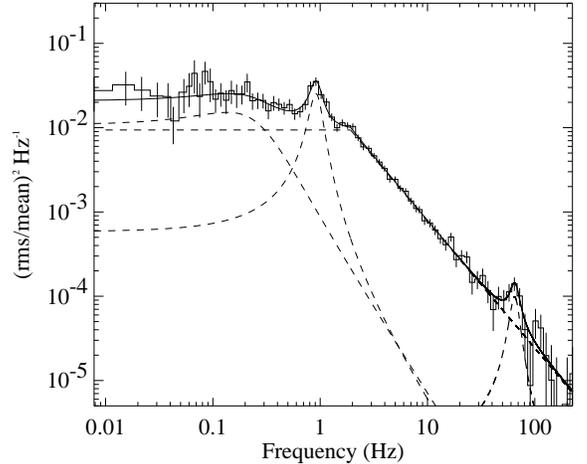}
\caption{\label{fig:ob4ps}  
Power spectrum for observation~4 with QPOs at 0.96 Hz and 65 Hz. The high 
frequency QPO has 4.9\% rms amplitude and a quality factor of 4.4. 
}
\end{figure}

     Perhaps the most important result is the detection of a \HQ\ in 
observation~4 at $\rm 65.0\,\pm\,2.5$ Hz (see Fig.~\ref{fig:ob4ps}). Within
the restricted energy band of 2 -- 28~keV, it has an $F$-test \citep{Bevington}
 value of 0.99997. For the whole energy band (2 -- 67 keV), it has an rms 
amplitude of 4.9\%, and a $Q$-value of 4.4. If we divide the light 
curve into three energy bands, the amplitude is $\rm 5.5\,\pm\,1.3$\% for 
2 -- 6~keV band and $\rm 4.7\,\pm\,1.1$\% for 6 -- 15~keV band. The QPO is not 
detected in the third energy band (15 -- 67 keV) with a 90\% confidence upper 
limit amplitude of 4.5\%. Thus, the strength of the \HQ\ is either constant or 
possibly decreasing in energy. 
 
\begin{figure}[t]
\plotone{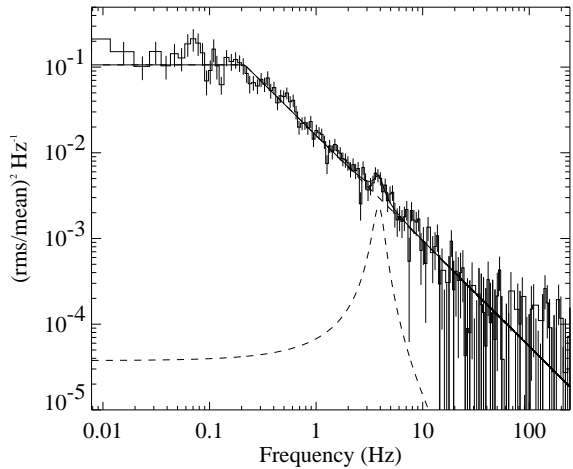}
\caption{\label{fig:ob9ps} 
Power spectrum for observation~9 with a probable detection of a QPO at 3.9 Hz. 
This QPO has 5.8\% rms amplitude and a quality value of $\rm \sim 4$.
}
\end{figure}

    We also report the probable detection of another QPO in observation~9 at 
$\rm 3.9\,\pm\,0.1$ Hz with an $F$-test value of 0.9997 (see 
Fig.~\ref{fig:ob9ps}). This QPO is well above the break frequency of the FT 
component in contrast to the low frequency QPOs that are observed in 
observations~1 -- 7. This is indicative of a different origin for the 
3.9~Hz QPO. In the 2 -- 67~keV energy band, it has a rms amplitude of 
$\rm 5.8\,\pm\,1.1\%$ and a $Q$-value of $\rm \sim 4$. When we divide the 
light curve into two energy bands, we observe that the amplitude of the QPO 
increases substantially from $\rm 4.6\,\pm\,1.7\%$ in 2 -- 8.5 keV to
$\rm 8.3\,\pm\,1.6\%$ in 8.5 -- 67 keV.  Although we do not associate this QPO 
with the \LQ s, it may be related to the 65~Hz QPO and we consider the 
implications of this possibility below.  However, we note that there are 
reports of QPOs above the break frequency that are probably not related to the 
\HQ s \citep{Remillard01, Homan00}.


\section{Discussion}

\subsection{The 65~Hz QPO}

     High frequency QPOs have been observed in five galactic BHCs. Two of
 them have fixed frequency over several observations, the 67 Hz QPO of 
GRS~1915+105 \citep{Morgan97} and 300~Hz QPO of GRO~J1655$-$40 
\citep{Remillard99_2}\footnote{Recently, pairs of high frequency QPOs were
 found in both sources \citep{Strohmayer01,Strohmayer01_b}.}. The 184 Hz QPO 
of 4U~1630$-$47 has been observed only once \citep{Remillard99_3}. 
XTE~J1859+226 \citep{Cui00} has variable \HQ s (150 -- 184 Hz ). Finally, 
\SO\ also has variable \HQ s with a frequency range of 100 -- 284 Hz 
\citep{Remillard99,Homan00,Miller01_a}. These \HQ s in \SO\ have rms 
amplitudes of around 1\% for the energy band of 2 -- 60 keV. Typically, the 
amplitudes increase with energy. The $Q$-values are between 4 and 13. They 
were observed in the very high state where the energy spectra show a strong 
power-law component \citep{Remillard99}. 

\begin{figure}[t]
\centerline{\includegraphics[height=5.5in]{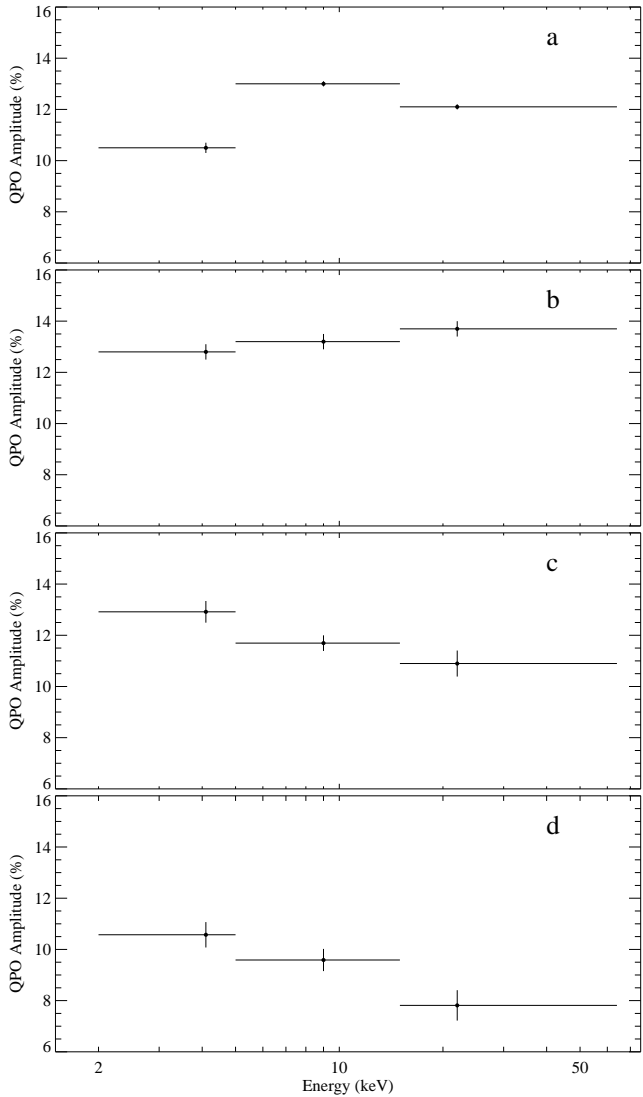}}
\vspace{0.5cm}
\caption{\label{fig:rmsan} 
Low frequency QPO amplitude as a function of energy. (a) observation 1, 
(b) observation 2, (c) observation 3, and (d) combination of observations 4
through 7. }
\end{figure}

 The 65~Hz QPO we observe is not in the previously observed frequency range
for \SO , but it is close to the the lower end of this frequency range
and also to the 67~Hz QPO observed for GRS~1915+105.  The $Q$-value for
the 65~Hz QPO is in the range usually seen for high frequency QPOs, but
the rms amplitude (4.9\%) is higher than typical values seen for high frequency
 QPOs.  Another difference is that the amplitude for the 65~Hz QPO does not 
increase with energy as is usually observed for high frequency QPOs.  It is 
possible that the differences reflect the fact that the 65~Hz QPO was observed 
during the transition to the hard state, while other high frequency QPOs have 
been detected when the systems were in the very high state. Although the 
properties of the 65~Hz QPO are not all typical of high frequency QPOs, the
 properties fit in well with trends that have been reported  by other authors. 
 For the high frequency QPOs in \SO , \cite{Homan00} find a relationship 
between the QPO frequency and the location of the source on a color-color 
diagram. Using the same energy bands that were used in that work, we have 
calculated the values of the soft color and the hard color for observation~4. 
 The soft color value of 0.91 and the hard color value of 0.08 are consistent 
with an extrapolation of the trend reported by \cite{Homan00}.  High and low 
frequency QPOs are often seen simultaneously in the power spectra of BHCs as 
we observe for \SO\ during observation 4. \cite{Psaltis99} have reported on 
correlations between the frequencies of the high and low frequency QPOs and/or
 noise components in the power spectra of BHC and neutron star systems.  The 
frequencies we measure for the observation~4 QPO pair are consistent with the
 \cite{Psaltis99} correlation that includes the high frequency QPOs for 
GRO~J1655$-$40 and previous observations of \SO. Overall, we believe that the 
evidence supports an association between the 65 Hz QPO and the previously 
observed QPOs.

    For a $\rm 7\; M_{\odot}$ non-rotating black hole, the dynamical 
time scale at the last stable orbit is 314 Hz. High frequency 
QPOs (65 -- 284 Hz) therefore arise close to the black hole
where effects of general relativity are expected to be significant. Several 
models have been proposed to explain these QPOs. A complete list of 
these models can be found in a review by \cite{Remillard00}. The three that can
 be applied to variable frequency QPOs are hot blobs in the inner edge of the 
disk (with Keplerian QPO frequency), Lense-Thirring precession of the tilted 
circular orbits of a spinning black hole (hereafter LT model) 
\citep{Cui98,Cui98_3}, and diskoseismic mode oscillations 
\citep{Nowak93,Nowak98}. To produce variable frequency QPOs, the first two 
require a change in the inner radius of the disk \citep{Cui00}, and the last 
can have variable frequency if the disk thickness ($h$) changes as a function 
of radius ($r$) \citep{Nowak98}. If the 65 Hz QPO in \SO\ is related to the 
284 Hz QPO observed previously, then any model must explain a factor of 4 
change in frequency. It was argued by \cite{Nowak93} that the change in 
frequency due to changes in ($h/r$) is small, and g-mode oscillations 
are expected to produce low amplitude ($\rm \sim\,1\%$) QPOs. The LT mechanism,
 on the other hand, can account for the change in frequency. It requires \HQ s 
to occur during state transitions because the accretion disk is expected to be 
realigned with the equatorial plane for stable accretion due to the 
Bardeen-Petterson effect and the QPO should disappear in time \citep{Cui98}. 
This is consistent with the occurrence of the 65~Hz QPO during a transition. 
However, it is not clear how the luminosity variations can be produced by LT 
precession \citep{Nowak98,Cui98}. Although hot blobs at the inner edge 
\citep[or in a region where disk emissivity peaks,][]{Cui98_3} can explain the 
high frequency QPOs and the change in the frequency, there are reports that 
this model is incompatible with the spectral results \citep{Cui98}.

     All the models described above have the \HQ\ originating in the accretion 
disk. \cite{Remillard99} observe that the \HQ s emerge in \SO\ when the
 disk temperature ($\rm T_{col}$) is high ($\rm >0.84\,keV$). In contrast, the 
disk-blackbody component is not detected in the \emph{RXTE} band for 
observation~4, indicating a disk temperature less than about 0.4 keV (paper I).
 This means that the power-law component provides nearly all of the 2 -- 150 
keV luminosity. The properties of previously observed QPOs seem to correlate 
much better with those of hard X-rays \citep[perhaps of a Comptonizing corona,]
[]{Cui98_2}. Our observations seem to support this trend since we observe a 
strong \HQ\ without a disk component in the spectrum. Note that this is not a 
proof that the origin of \HQ s cannot be in the disk since the oscillations in 
the disk flux (soft X-rays and UV) can modulate the input to the corona. 

    As discussed above, the high QPO frequencies may be associated with the 
Keplerian or the LT time scale for the inner edge of the accretion disk. Under 
this assumption, one can use the QPO frequencies to trace the radius of the 
inner edge of the accretion disk. The highest frequency observed in \SO\ is 
284 Hz, which likely originates close to the black hole. We designate 
$\rm R_{min}$ to be the inner radius when the 284 Hz \HQ\ is observed. 
This radius may be close to the marginally stable orbit. If we assume that the 
65~Hz QPO in observation~4 and the 3.9~Hz QPO in observation~9 are related to 
the 284~Hz QPO, then, for Keplerian time scales, the inner edge moves to 
$\rm 2.7\,R_{min}$ for observation~4 and $\rm 17\,R_{min}$ for observation~9. 
 The change in radius for a LT time scale is smaller, $\rm 1.6\,R_{min}$ and 
$\rm 4.2\,R_{min}$, respectively. 

    One of the models devised to understand the energy spectra of accreting 
BHCs is the ADAF (advection-dominated accretion flow) model \citep{Narayan96}. 
\cite{Esin97} developed a model consisting of an ADAF and an optically thick 
outer accretion disk and used this model to explain the spectral changes seen 
in the BHC Nova Muscae. In the context of this model, the different spectral 
states can be explained by changes in the mass accretion rate and the inner 
edge of the optically thick accretion disk.  This model predicts that the 
inner edge of the accretion disk moves away from the black hole during a 
transition to the hard state, which is consistent with the increasing of the 
inner radius discussed previously. If the association of the \HQ s to the 
Keplerian time scales at the inner edge of the disk is correct, then the 
inner edge of the disk moves substantially even in the intermediate/very high 
state where the QPOs have been observed in 100 -- 284 Hz range.

\subsection{The low frequency QPOs and the continuum parameters during the 
state transition}
  
  Observations by \cite{Cui99} in the beginning of the 1998 outburst 
and the observations reported here demonstrate the behavior of the temporal 
properties of \SO\ during the state transitions between the hard and the very 
high state. As the flux decreases and the source makes its transition to the 
hard state, the QPO frequency decreases along with the break frequency, while 
the amplitude of the FT component and the index of the FT power-law component 
increase.  The reverse was seen by \cite{Cui99} when the source made its 
transition to the very high state. The temporal properties for 
\SO\ during the transition to the hard state are very similar to what has been 
observed for the BHC 4U~1630$-$47 \citep{Tomsick00}.  For 4U~1630$-$47, a QPO 
was present that dropped in frequency from 3.4 to 0.2~Hz as the source flux 
decayed.  Also, the evolution of the continuum parameters were similar with an 
increase in the amplitude of the FT component and a drop in the break 
frequency. However, for 4U~1630$-$47, a sharp transition was seen with most of 
the changes in the temporal properties occurring in less than 2 days.  For 
\SO , the QPO was present and the FT rms amplitude was relatively high in our 
first observation. It is possible that a sharp transition might have happened 
before our observations began.  One interesting feature generally over-looked 
in describing the state transitions is the change in the index of the power-law
 component in the power spectrum. This regular change of index during state 
transitions is observed in \SO\ and also in GRO~J1655$-$40 \citep{Mendez98} and
may have implications for the models of aperiodic variability such as shot 
noise models \citep[][and references therein]{Pottschmidt98}.

    The \SO\ power spectra show a correlation between the break frequency and 
the central frequency of the \LQ . The fact that the break frequency follows 
the QPO frequency was reported by \cite{Cui99} for \SO\, and this trend is 
observed in other black holes and  neutron star systems \citep{Wijnands99}. 
Fig.~\ref{fig:vbvf} shows this correlation for two cases; crosses indicate the
 cases where we do not include the first harmonic in the fit and circles 
indicate the cases where we include a first harmonic (even if it is not 
statistically significant). Except for observation 1, including the 
harmonic does not alter the break frequencies significantly. For observation 
1, the fit without the harmonic yields a break frequency of 8.6 Hz, compared to
5.0 Hz with the harmonic.  Some scatter is present in the correlations 
reported by  \cite{Wijnands99}, and they attribute this to the complex 
structure of the QPO and its harmonics. Similarly, our correlation breaks down 
in observation~1 when we include the first harmonic in the fit (circles in 
Fig.~\ref{fig:vbvf}). If we do not include the first harmonic, there is no 
scatter and the correlation is superb with a correlation coefficient of 0.9996.
 But the fact that the correlation breaks down for the full model and also 
that the break frequencies are almost twice  the QPO frequencies lead us to
 believe that the first harmonic, drives the break frequency to the centroid 
frequency of the first harmonic.

    Previous results by \cite{Cui99}, \cite{Homan00} and \cite{Remillard99} 
show that the amplitude of the both low and \HQ s increase with energy in \SO. 
For some cases there is a turnover in this relation, but it is at high 
energies ($\rm >\,60\,keV$) as shown by \cite{KalemciHEAD00}. This kind 
of turnover is observed also in GRS~1915+105 \citep{Tomsick01}. In contrast,
we observed a turnover in observation~1 at 10 keV and a decreasing relation
after observation~3 (see Fig.~\ref{fig:rmsan}). A decrease in the amplitude as
 energy increases was also seen in the hard state of Cyg X-1 
\citep{Rutledge99}, and for some low frequency QPOs (0.06 - 0.2 Hz) in 
GRS 1915+105 \citep{Morgan97}. A detailed investigation of the amplitude vs. 
energy relation in \SO\ is subject of a future work. 

\section{Summary and Conclusions}

    The most important result of this paper is the detection of a \HQ\ 
at 65 Hz since it is the first time a \HQ\ has been observed in a transition to
 the hard state during the outburst decay. It has relatively high amplitude of 
4.9\% and a $Q$-value of 4.4. During the time the QPO was observed, the 
\emph{RXTE} energy spectrum was dominated by a power-law component and the 
disk component was not detected (paper I). Although the evidence supports 
its association with the previously observed \HQ s, it is stronger in 
amplitude and does not show an increasing amplitude with increasing energy. We
 considered three mechanisms to explain the variable \HQ s: self-illuminating 
clumps in the inner accretion disk; Lense-Thirring precession; and 
diskoseismic g-mode oscillations. The diskoseismic oscillation model is 
unlikely to explain such a high amplitude. We also report the probable 
detection of a QPO at 3.9~Hz during an observation that occurred after the one
 where 65~Hz QPO was detected. This QPO is well above the break frequency, and 
it is unlikely that it is related to the other low frequency QPOs.  It is 
possible that the 3.9~Hz and 65~Hz QPOs are related, and that they provide 
information about the location of the inner radius of the accretion disk. 

     Another result is the correlation between the \LQ s and the break 
frequencies and the fact that it breaks down when a complex QPO structure is 
present. We also show that the change in temporal properties in the transition
 to the hard state is the mirror image of the change in the temporal properties
 in hard to very high state transition \citep{Cui99}. The main features of the 
transition are: a drop in the \LQ\ frequency from 4.1 Hz to 0.36 Hz, a decrease
 in the break frequency from 5 Hz to 0.6 Hz, an increase in FT amplitude from 
14\% to 42\% and an increase of the power-law index of the power spectra from 
$-$1.8 to $-$1.1 (see Table~\ref{table:par}). From this data, it is not clear 
if a sharp transition exists in temporal properties as reported for 
4U~1630$-$47 \citep{Tomsick00}. We should be able to conclude whether a sharp 
transition exists by combining our data with the data taken prior to our 
observations.
 

\acknowledgments 
EK acknowledges useful discussions with J\"{o}rn Wilms and Wayne Coburn. The
authors would like to thank all scientists contributed to the T\"{u}bingen 
Timing Tools. In part, this material is based upon work supported by the 
National Aeronautics and Space Administration under grants  NAG5-30120 and  
NAG5-10886. EK was partially supported by T\"{U}BITAK. KP is supported by DFG 
grant Sta 173/25-3 and a travel grant from the DAAD and would like to thank 
CASS for its hospitality. PK acknowledges partial support from NASA grant 
NAG5-7405.






\begin{table}[ht]
\caption{\label{table:par} Power Spectra Fit Parameters}
\begin{minipage}{\linewidth}
\scriptsize
\begin{tabular}{c|c|c|c|c|c|c|c|c|c|c} \hline \hline
 MJD\footnote{Modified Julian Date ($\rm MJD =JD-2400000.5$) at the midpoint 
of the observation.} & Obs. & Flux\footnote{Flux in 2.5--20 keV range, in 
units of $\rm 10^{-9}\,ergs\,cm^{-2}\,s^{-1}$.} & FT rms  & FT index & 
FT break & LF\footnote{Low frequency noise component.} freq. & LF rms  & 
HF\footnote{High frequency noise component.} freq. & HF rms &
 $\rm \chi^{2}/\nu$ \\ 
   &   &  & (\%) &   & (Hz) & (Hz)  & (\%) & (Hz) & (\%) & \\ \hline
 51680.4  & 1  & 5.62 & $\rm16.4\pm0.1$ & $\rm-1.70\pm0.05$ & $\rm5.04\pm0.05$ 

& $\rm 0.73\pm0.03$ & $\rm10.1\pm0.3$ & $\rm 62.9\pm30.5$ & $\rm8.7\pm1.4$ & 
153.7/144 \\

 51682.3  & 2  & 4.49 & $\rm20.4\pm1.2$ & $\rm-1.82\pm0.08$ & $\rm4.87\pm0.13$ 
 
& $\rm 0.39\pm0.03$ & $\rm12.8\pm0.5$ & $\rm 49.0\pm145.9$ & $\rm11.8\pm6.8$ & 148.7/144 \\ 

 51683.8  & 3  & 3.69 & $\rm23.8\pm1.3$ & $\rm-1.56\pm0.04$ & $\rm2.41\pm0.07$ 

& $\rm 0.26\pm0.01$ & $\rm9.1\pm0.4$ & $\rm 40.0\pm26.6$ & $\rm7.1\pm2.5$ & 152.4/144 \\

 51684.8  & 4\footnote{The \HQ\ at 65 Hz is detected in this observation.} & 3.20 & $\rm23.9\pm2.3$ & $\rm-1.48\pm0.03$ & $\rm1.95\pm0.12$ 

& $\rm 0.10\pm0.09$ & $\rm11.1\pm2.4$ & - & - & 110.7/139 \\

 51686.3  & 5  & 2.68 & $\rm24.8\pm1.8$ & $\rm-1.39\pm0.02$ & $\rm1.46\pm0.07$ 

& $\rm 0.23\pm0.01$ & $\rm7.6\pm0.6$ & - & - & 145.5/147 \\

 51687.2  & 6  & 2.38 & $\rm27.1\pm3.2$ & $\rm-1.32\pm0.03$ & $\rm1.18\pm0.09$ 

& $\rm 0.16\pm0.02$  & $\rm8.7\pm1.2$ & - & - & 123.1/130 \\

 51688.8  & 7  & 1.85 & $27.0\pm4.0$ & $\rm-1.33\pm0.04$ & $\rm0.93\pm0.07$ 

& $\rm 0.17\pm0.04$ & $\rm16.5\pm4.2$ & - & - & 130.5/130 \\ 

 51690.8  & 8 & 1.43 & $\rm28.7\pm4.9$ & $\rm-1.25\pm0.04$ & $\rm0.81\pm0.07$ 

& $\rm 0.14\pm0.02$ & $\rm18.8\pm1.7$ & -  & - & 192.0/133 \\ 

 51692.6  & 9\footnote{The 3.9 Hz QPO is detected in this observation.}  & 1.04 & $\rm34.7\pm3.9$  & $\rm-1.23\pm0.03$ & $\rm0.22\pm0.01$ & - & -  &  -  & -  & 161.4/133 \\ 
 51693.4  & 10 & 0.91 & $\rm34.4\pm2.1$  & $\rm-1.18\pm0.02$ & $\rm0.16\pm0.01$ & - & -  &  - & - & 128.1/136 \\ 
 51695.3  & 11 & 0.72 & $\rm40.8\pm8.3$  & $\rm-1.13\pm0.03$ & $\rm0.14\pm0.01$ & - & -  &  - & - & 119.4/136 \\ 
 51696.5  & 12 & 0.64 & $\rm42.5\pm7.3$  & $\rm-1.12\pm0.02$ & $\rm0.09\pm0.01$ & - & -  &  - & - & 135.0/136 \\ 
 51698.9  & 13 & 0.46 & $\rm65.8\pm48.8$  & $\rm-1.05\pm0.05$ & $\rm0.05\pm0.01$ & - & -  &  - & - & 64.6/91 \\ \hline
\end{tabular}
\end{minipage}
\\

\end{table}

\begin{table}[ht]
\caption{\label{table:qpo} Low frequency QPO fit parameters}
\begin{minipage}{\linewidth}
\scriptsize
\begin{tabular}{c|c|c|c|c|c} \hline \hline
 Obs. & QPO freq. & QPO rms & FWHM & harm. freq.\footnote{Frequency of the 
first harmonic.} & harm. RMS  \\ 
    & (Hz) & (\%) & (Hz) & (Hz) & (\%) \\ \hline
1 & $\rm 4.090\pm0.015$ & $\rm 10.4\pm0.2$ & $\rm 0.72\pm0.04$ & $\rm 8.75\pm0.15$ & $\rm 5.8\pm0.4$ \\
2 & $\rm 2.337\pm0.010$ & $\rm 12.8\pm0.3$ & $\rm 0.44\pm0.03$ & $\rm 4.77\pm0.04$ & $\rm <1.8$\footnote{90\% confidence upper limit.} \\
3 & $\rm 1.119\pm0.007$ & $\rm 11.9\pm0.3$ & $\rm 0.26\pm0.02$ & $\rm 2.35\pm0.02$ & $\rm <2.0^{\it{b}}$\\
4 & $\rm 0.962\pm0.016$ & $\rm 10.0\pm0.8$ & $\rm 0.28\pm0.05$ & $\rm 1.83\pm0.05$ & $\rm <4.2^{\it{b}}$\\
5 & $\rm 0.678\pm0.010$ & $\rm 10.7\pm0.6$ & $\rm 0.19\pm0.03$ & - & -\\
6 & $\rm 0.559\pm0.012$ & $\rm 11.4\pm1.0$ & $\rm 0.21\pm0.05$ & - & -\\
7 & $\rm 0.355\pm0.005$ & $\rm 7.1\pm0.9$ & $\rm 0.05\pm0.02$ & - & -\\ 
8 & 0.01 -- 0.8 \footnote{Frequency range searched (up to the break frequency).}& $\rm <7.3^{\it{b}} $ & $\rm 0.15 \times freq.$  \footnote{The width is fixed to 0.15 $\rm \times$ QPO frequency ($Q$ = 6.6).} & - & - \\ \hline 
\end{tabular}
\end{minipage}
\end{table}


\end{document}